\documentclass[12pt]{article}
\usepackage{amsmath,amssymb,amsfonts,amsthm,array,tikz,authblk}
\usepackage{graphicx,wrapfig,parskip,enumitem}
\usepackage{caption,graphicx,wrapfig,parskip,enumitem,multirow,lscape}
\usepackage[T1]{fontenc}
\usepackage[left=1in, right=1in, top=1in, bottom=1in]{geometry}
\usepackage{etoolbox}
\usepackage{titlesec}
\usepackage{url}
\usepackage{verbatim}
\usepackage{lineno}

\AtBeginEnvironment{quote}{\vspace{-\topsep}\small}
\AtEndEnvironment{quote}{\vspace{-\topsep}}

\title{Pattern detection in bipartite networks:\\a review of terminology, applications and methods}
\date{\vspace{-5ex}}
\author[1]{Zachary Neal*}
\author[2]{Annabel Cadieux}
\author[3,4,5]{Diego Garlaschelli}
\author[6]{Nicholas J. Gotelli}
\author[7]{Fabio Saracco}
\author[4,5]{Tiziano Squartini}
\author[8]{Shade T. Shutters}
\author[9]{Werner Ulrich}
\author[10]{Guanyang Wang}
\author[11]{Giovanni Strona*}

\affil[1]{Michigan State University, US}
\affil[2]{Mathedoc, <https://mathedoc.de/>, Germany}
\affil[3]{Lorentz Institute for Theoretical Physics, the Netherlands}
\affil[4]{IMT School for Advanced Studies, Italy}
\affil[5]{INdAM, Italy}
\affil[6]{University of Vermont, US}
\affil[7]{Enrico Fermi Research Center, Italy}
\affil[8]{Arizona State University, US}
\affil[9]{Faculty of Biological and Veterinary Sciences at Nicolaus Copernicus University, Poland}
\affil[10]{Rutgers University, US}
\affil[11]{European Commission, Joint Research Centre, Italy}

\begin{document}
	\maketitle
	*Correspondence to: Z.N. (\url{zpneal@msu.edu}) and G.S. (\url{giovanni.strona@ec.europa.eu})

\newpage
\section*{Abstract}
Two dimensional matrices with binary (0/1) entries are a common data structure in many research fields. Examples include ecology, economics, mathematics, physics, psychometrics and others.  Because the columns and rows of these matrices represent distinct entities, they can equivalently be expressed as a pair of bipartite networks that are linked by projection. A variety of diversity statistics and network metrics can then be used to quantify patterns in these matrices and networks. But what should these patterns be compared to? In all of these disciplines, researchers have recognized the necessity of comparing an empirical matrix to a benchmark set of "null" matrices created by randomizing certain elements of the original data. This common need has nevertheless promoted the independent development of methodologies by researchers who come from different backgrounds and use different terminology. Here, we provide a multidisciplinary review of randomization techniques for matrices representing binary, bipartite networks. We aim to translate the concepts from different technical domains into a common language that is accessible to a broad scientific audience. Specifically, after briefly reviewing examples of binary matrix structures across different fields, we introduce the major approaches and common strategies for randomizing these matrices. We then explore the details of and performance of specific techniques, and discuss their limitations and computational challenges. In particular, we focus on the conceptual importance and implementation of structural constraints on the randomization, such as preserving row or columns sums of the original matrix in each of the randomized matrices. Our review serves both as a guide for empiricists in different disciplines, as well as a reference point for researchers working on theoretical and methodological developments in matrix randomization methods.

\section{Introduction}
\begin{figure}[!htbp] 
	\vspace{20pt}
	\begin{center}
		\includegraphics[width=0.8\textwidth]{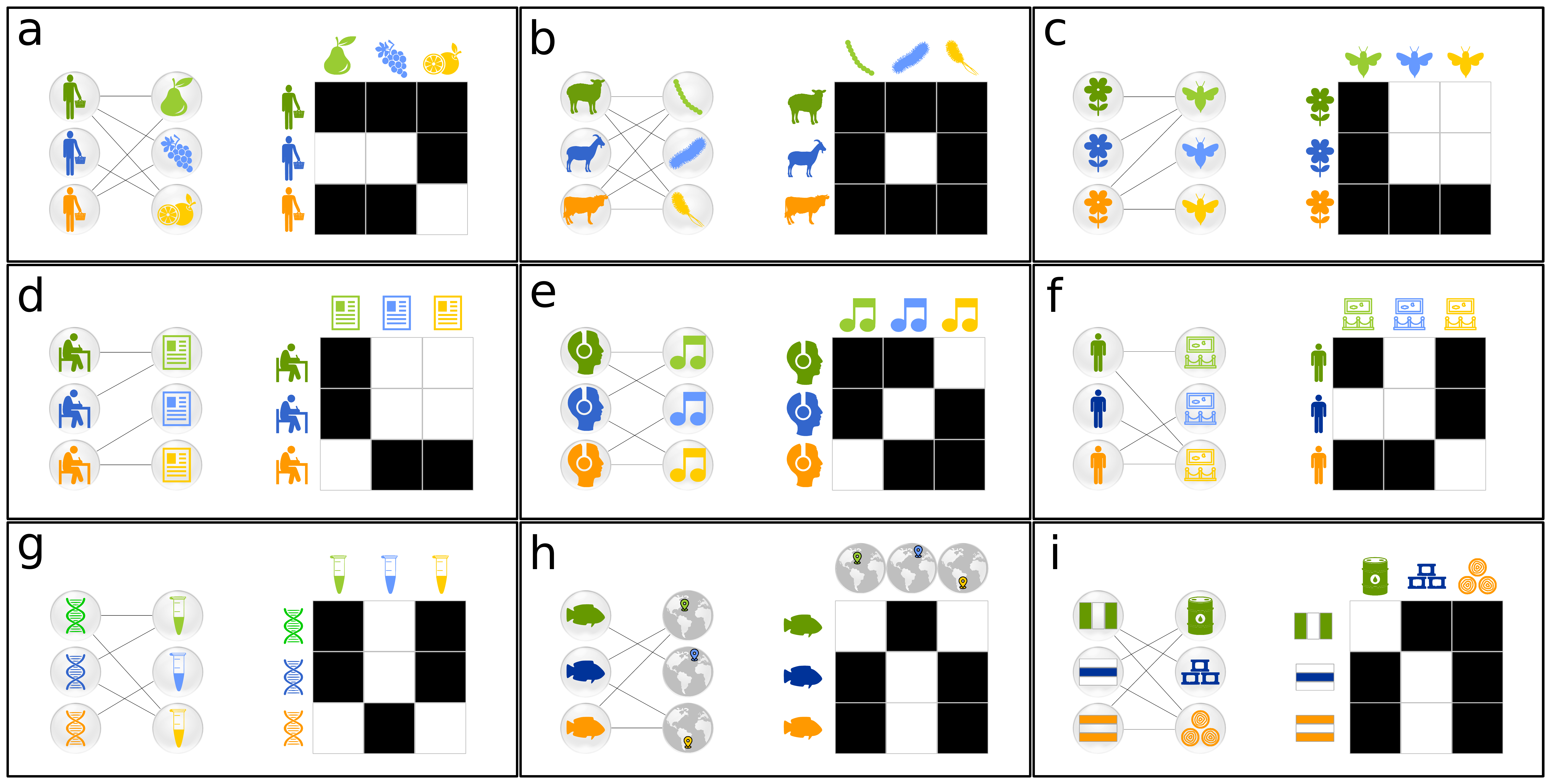}
	\end{center}
	\caption{\textbf{Bipartite networks are ubiquitous in the real world.} Examples of systems that can be represented by bipartite networks (left)and their corresponding binary matrix representation (right). Black/grey cells in each matrix indicate presence of links (i.e. $1$s) between the items in rows and the items in columns, while white cells indicate the absence of links (i.e. $0$s). The networks link, respectively: (\textbf{a}) buyers to purchases; (\textbf{b}) ruminants to associated microbiota; (\textbf{c}) plants to pollinators; (\textbf{d}) authors to articles; (\textbf{e}) listeners to songs; (\textbf{f}) visitors to exhibitions; (\textbf{g}) genes to samples; (\textbf{h}) species occurrences to localities; (\textbf{i}) countries to exported commodities.}
	\label{fig1} 
\end{figure}

What do ecological metacommunities, biotic interactions, gene mutations, international trade, public transportation, musical preferences and organized group events have in common? They all represent systems that we can conveniently synthesize and investigate as bipartite networks~\cite{gotelli2000null,kim2016somatic,saracco2015randomizing,chen2007study,lambiotte2005uncovering,smiljanic2017associative}. Indeed, some have argued that ``any complex network [i.e. system] may be viewed as a bipartite graph'' ~\cite{guillaume2006bipartite}. These graphs provide information about  the presence or absence of relationships between two entities or of the strength of these relationships. Accordingly, numerous examples of application of bipartite networks can be found across multiple research fields such as anthropology, biology, ecology, economics, engineering, finance, logistics, management, mathematics, physics, social sciences (see fig.~\ref{fig1} and table~\ref{table1}). The present review deals with presence – absence networks only.

\begin{figure}[!htbp]
	\vspace{20pt}
	\begin{center}
		\includegraphics[width=0.7\textwidth]{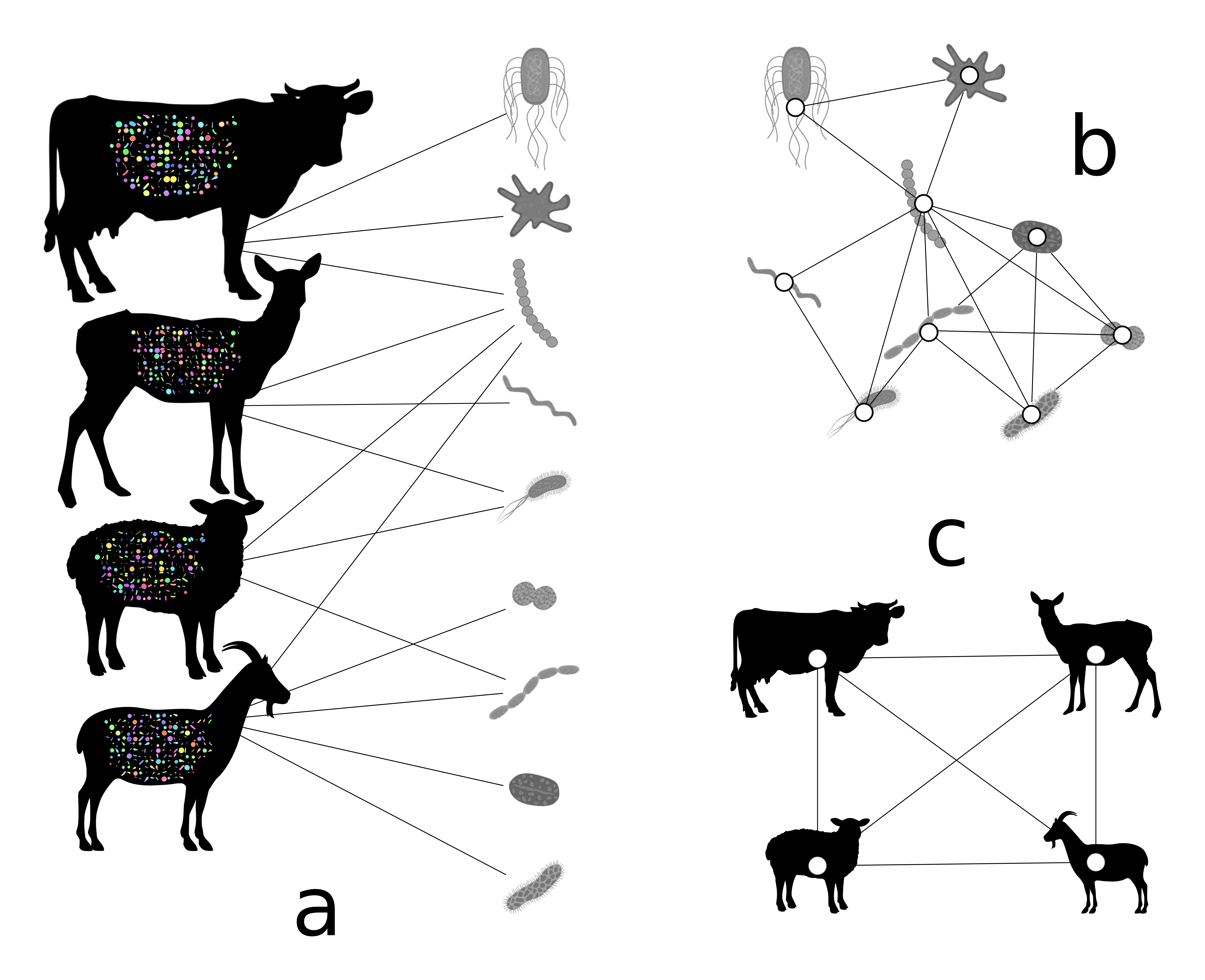}
	\end{center}
	\caption{\textbf{Bipartite networks and their projections.} A hypothetical bipartite network connecting different ruminants to associated microorganisms (\textbf{a}), and its two one-mode projections. One projection (\textbf{b}) connects all the microorganisms that are found together in at least one host. The other projection (\textbf{c}) connects all the ruminants sharing at least one microorganism, generating a fully connected network in this example.}
	\label{fig2} 
\end{figure}

Many contexts benefit from directly analyzing bipartite networks. However, when a bipartite network's two disjoint sets of nodes represent distinct types of entities (as they do in fig.~\ref{fig1}), it is also known as a two-mode network, which can also be transformed via projection into a unipartite one-mode network. In such a two-mode projection, pairs from one set of nodes are connected when they share links to nodes from the other set. As with bipartite networks, analysis of two-mode projections is commonly used in multiple fields. For instance, in bibliometrics, a bipartite network connecting scientific publications to their authors can be projected onto a co-authorship network synthesizing scientific collaborations~\cite{newman2004coauthorship}. Other common examples of bipartite network projections include those mapping legislative collaboration through bill co-sponsorship in political science~\cite{neal2020sign}, gene interactions through sample co-expression in genetics~\cite{zhang2005general} and bacterial interactions through sample co-occurrence in microbial ecology ~\cite{freilich2010large} (see fig.~\ref{fig2}). Actually, any matrix/network mapping the simultaneous presence (i.e. co-occurrence) of items, organisms, or events in space and/or time can be considered a bipartite network projection of the ideal bipartite network linking items/organisms/events to different localities and/or times~\cite{filho2020,guillaume2004bipartite,newman2003social}.

The ubiquity of bipartite networks and their projections has resulted in a considerable amount of theoretical and applied knowledge. On the other hand, their vast interdisciplinary span has prevented the convergence of knowledge into an organic corpus. One critical task where this lack of convergence presents significant barriers is the detection of patterns in bipartite networks~\cite{strona2018bi}. Different research fields have developed their own (sometimes duplicate) methods for this task, using field-specific terminology and summarizing their findings in field-specific reviews (e.g. in ecology~\cite{gotelli2000null},  social~\cite{neal2014backbone} and computer science~\cite{zweig2011systematic} and complex systems\cite{cimini2019statistical}). For example, one widely used method is known as ``stochastic degree sequence model'' in social science~\cite{neal2014backbone}, but as a ``canonical configuration model'' in statistical physics~\cite{Squartini2011a}. Despite the widespread applicability of these methods, their strong intra-disciplinary focus has hindered progress, often forcing individual fields to rediscover methods without fully benefiting from innovations in other areas. 

To confront and overcome these challenges, this review provides a comprehensive, multidisciplinary synthesis of available knowledge on bipartite network randomization techniques, with the ultimate aim to enhance progress and prevent the wheel from being reinvented multiple times. To tackle these aims in a way that will be accessible to a broad audience, we provide conceptual illustrations and concrete examples to identify overlaps and differences between the approaches developed in the various fields of science, limiting the use of formal notation to specific technical sections dedicated to readers interested in the formal specifications of the models we discuss.

\section{What are bipartite networks?}
As with any network, a bipartite network is composed of a set of \textit{nodes}, pairs of which may be connected by \textit{edges}. The essential feature of a bipartite network is that these nodes can be partitioned into two sets such that edges exist only between these sets. This feature is clear in each panel of fig.~\ref{fig1}, where edges connect nodes in the left row (e.g., panel 1, buyers) to nodes in the right row (e.g., panel 1, purchases), but never to nodes in the same row.

A bipartite network can be represented as a graph, where nodes are drawn as dots and edges are drawn as lines connecting them (left half of each panel in fig.~\ref{fig1}). It can also be represented as a binary matrix where rows correspond to one set of nodes, columns correspond to the other set of nodes, and entries contain a 1 if the row-node is connected to the column-node (right half of each panel in fig.~\ref{fig1}). Graph representations are often useful for visualization, while matrix representations are more useful for formal analysis.

Whether represented as a graph or matrix, bipartite networks can be characterized by several properties. Here we focus on two that play a particularly important role in the pattern detection methods we discuss below. First, a bipartite network can be characterized by its \textit{density}, which is the fraction of possible edges that are present. In its matrix representation, the density is simply the proportion of filled cells (i.e. cells with entry 1). For example, the density of the bipartite network shown in fig.~\ref{fig1}a is .66 because 6 of a possible 9 edges are present. Second, a bipartite network can be characterized by its nodes' degree sequences, which capture each node's number of connections. In its matrix representation, the degree sequences are given by the matrix's row and column sums. For example, the degree sequence for the row nodes in fig.~\ref{fig1}a is \{3,1,2\}, while the degree sequence for its column nodes is \{2,2,2\}.

The defining feature of a bipartite network compared to a general network is its partitionability into two sets of nodes which are only connected in between but not inside the sets. However, in real world bipartite networks, those two sets of nodes often represent distinctly different types of entities, in which case the bipartite network is also called a two-mode network. For example, the nodes in the bipartite network shown in fig.~\ref{fig2}a represent distinctly different types of entities: ruminants on the left, and microorganisms on the right. A two-mode network can be analyzed as a bipartite network, however it can also be transformed into two, one-mode unipartite networks via projection, each consisting of the nodes of only one mode. An edge between two nodes in a projection exists if and only if these nodes are connected to at least one common node of the other mode in the two-mode network. For example, fig.~\ref{fig2}c illustrates that the cow and deer are connected in a one-mode projection because they are both connected to the same long worm-shaped microorganism in the two-mode network. The matrix representation of each one-mode projection (e.g. the one linking microorganisms occurring in the same ruminant, fig.~\ref{fig2}b; and the one linking ruminants sharing microorganisms, fig.~\ref{fig2}c) is the product of the two-mode network's matrix representation with its transpose and vice versa.

\section{Where are bipartite networks found?}
One primary focus in the analysis of bipartite networks and one-mode projections is pattern detection. Before turning to the different methods, for the sake of concreteness, we briefly review the range of contexts where such networks and patterns are observed.

\subsection{Ecology and Biogeography}
Employing null models of binary matrices for pattern detection has a long history in ecology \cite{gotelli2012statistical} and became particularly important in the context of the ongoing debate on how species interactions, particularly competition, determine the spatial distribution of species \cite{gotelli2002species,ulrich2004species,stone1990checkerboard,gilpin1982factors}. In this context, the distribution of species across a set of localities is represented as a bipartite network where the species (one set of nodes) are connected to the localities (the other set of nodes) where they are found at a given point in time. Comparing an observed bipartite network of species location to a set of randomized versions of the same network has allowed researchers to investigate structural patterns in both species-locality matrices and bipartite plant-pollinator ecological networks~\cite{patterson1986nested,bascompte2003nested,Borras2019,Bruno2020,caruso2021fluctuating}. Additionally, a bipartite species location network can be projected into a unipartite species co-location network, where similar methods allow researchers to evaluate whether pairs of species are consistently found in the same locations \cite{gotelli2000null}. This topic has received a lot of interest in recent years. In fact, it was at the center of a lively scientific debate on the possibility of inferring biotic interactions (and hence deriving ecological networks of interacting species) from species co-occurrence data \cite{morales2015inferring,blanchet2020co}, based on the idea that ecological interactions might play a fundamental role in determining overlapping (or segregated) species distribution patterns.

\subsection{Social Sciences}
In the social sciences, bipartite networks are frequently used to represent individuals' (one set of nodes) affiliations with objects (the other set of nodes). For example, in a sociological context they can represent individuals' membership in clubs~\cite{breiger1974duality}, while in a political science context they can represent legislators sponsorship of bills~\cite{neal2014backbone}. Limited extensions of classical network analytic techniques make it possible to describe and analyze patterns in social bipartite networks~\cite{faust1997centrality,wang2013exponential}, however it is more common for social scientists to study one-mode projections. The one-mode projection of a person-club bipartite network yields a network of individuals connected by their club co-memberships, while the one-mode projection of a legislator-bill bipartite network yields a network of legislators connected by their bill co-sponsorships. Pattern detection methods can be employed to determine when a dyad's number of co-memberships or co-sponsorships exceeds what would be expected by chance, and therefore can be treated as a proxy for an unobserved relationship of interest such as friendship or collaboration.

\subsection{Psychology}
In psychology, data capturing individuals' responses to psychological survey items can be represented as a bipartite network. In such a network, the respondents serve as one set of nodes, while the items serve as the other set of nodes, with each respondent nodes connected to each item nodes by an edge weighted with the response, often arbitrary ordinal (Likert) values. In psychometrics, such data is frequently analyzed using a dichotomous Rasch model to construct and score educational and psychological tests~\cite{rasch1960studies,rasch1993probabilistic} . Estimation of a Rasch model involves identifying patterns in the bipartite network by comparing it to a series of randomized alternatives~\cite{verhelst2008efficient}. More recently, psychologists have also explored the use of one-mode projections of bipartite clinical data, generating networks of symptom co-occurrence or co-morbidity~\cite{borsboom2021network}, which requires determining when such co-occurrence patterns exceed what would be expected by chance and relies on methods that remain subject to debate~\cite{neal2022critiques}.

\subsection{Economics}
Bipartite networks have found broad application in economics where they can represent products produced by countries \cite{hidalgo2007,Hidalgo2009,Hausmann2011,Tacchella2012,Caldarelli2012a,Cristelli2013, Cristelli2015}, skills required by occupations \cite{alabdulkareem2018,kok2014cities}, location of industries in cities \cite{neffke2011regions,oclery2019}, occupations \cite{galetti2022types,muneepeerakul2013urban}, or patent technologies \cite{toth2022technology,oneale2021structure}. Analysis of such data often focuses on the one-mode projection of these networks. For example, a growing branch of research known as ``Economic Complexity'' has recently focused on identifying the productive capabilities of the various countries through the analysis of their exported products~\cite{Hausmann2011,Tacchella2012}. As in other cases, such analysis rests on determining when patterns in countries' exports of a goods exceed random levels, taking into account such factors as the good's rarity.

\section{How are patterns detected in bipartite networks?}

\begin{figure}[!htbp]
	\vspace{20pt}
	\begin{center}
		\includegraphics[width=0.9\textwidth]{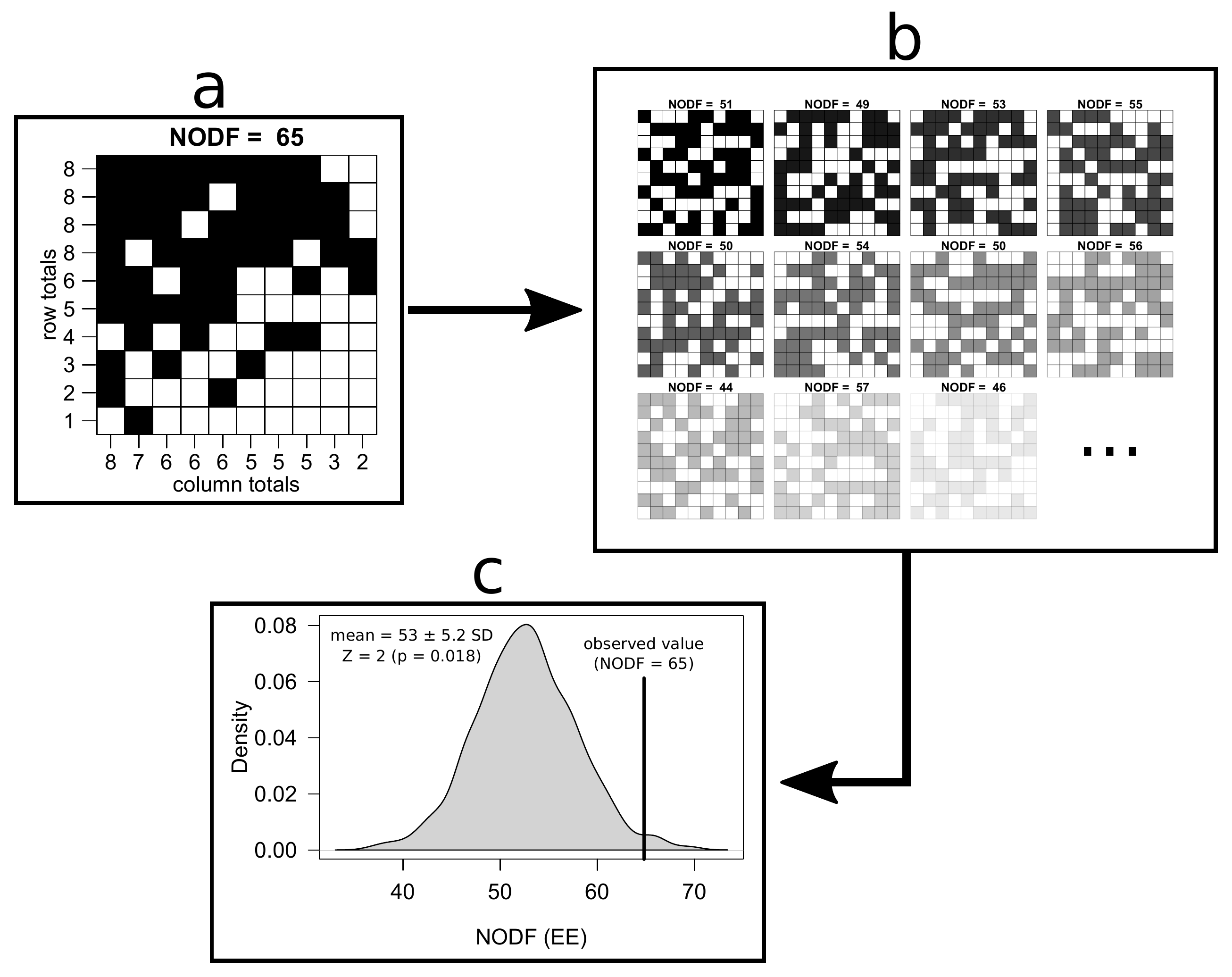}
	\end{center}
	\caption{\textbf{Pattern detection in bipartite networks.} Schematic example of how matrix randomization is used to detect non-random patterns in bipartite networks/rectangular matrices. Black/grey cells in each matrix indicate presence of links (i.e. $1$s) between the items in rows and the items in columns, while white cells indicate the absence of links (i.e. $0$s). First, the structural measure of interest (in this case a nestedness metric, NODF \cite{almeida2008consistent}) is computed on the target matrix (a). Then, a large set (ideally some hundreds or thousands) of randomized versions of the starting matrix are generated, and the target metric is computed for each of them (b). The possible rules to be applied in the generation of the random matrices, the reasoning behind and the implications of choosing a particular set of rules over another, as well as the practical implementation of randomization procedures will be described in detail in the sections. The target metric computed on the original matrix will be then compared with the distribution of the metric values computed on the random matrices. Such comparison would permit to obtain an estimated $p$-value computed as the frequency of random matrices for which the target metric is equal or higher than that of the original matrix. In some fields, and particularly in ecology \cite{strona2014methods}, it is also common practice to compute a standardized effect size ($Z$) as $(\mu - x)/\sigma$, where $\mu$ and $\sigma$ are the average and standard deviation of the target metric across the randomized matrices, and $x$ is the value of the metric in the original matrix. It should be noted that the use of $Z$ values is based on the underlying assumption that the distribution of the target metric values in the set of randomized matrices follows a normal distribution, which might not be always the case.}
	\label{fig_pattern_analysis} 
\end{figure}

Detecting non-random patterns in bipartite networks (i.e. patterns which are not likely to be seen by chance in a set of networks with certain properties) follows a fairly simple procedure (Fig.~\ref{fig_pattern_analysis}). First, a statistic of interest is computed from an observed bipartite network. The specific statistic depends entirely on the substantive research question. For example, in ecology, the compositional change among communities ($\beta$-diversity) has been quantified by a dozen of different measures, each focusing on different aspects of change \cite{tuomisto2010diversity}. In political science, one may explore the structure of cosponsorship networks in terms of number of bills co-sponsored by two legislators~\cite{neal2014backbone}.

Second, a random network is generated. For that, the observed bipartite network is randomized (we discuss how in section~\ref{sec:how}) in a way the preserves certain features of the original network (we discuss which ones in section~\ref{sec:what}). This new, random bipartite network arises from a `null model', so called because any patterns present in the original should be nullified by the randomization process.
Those features, technically the constraints of the null model, in fact define how a random pattern should statistically look like in the given context. For a given situation prior work and experience provides an expectation about the pattern of interest, and consequently also about the complementary null expectation. This expectation restricts the null space and our null model has to account for these constraints. The set of random bipartite networks is known as an ensemble, and each randomized bipartite network is mathematically a sample which is randomly drawn with (approximately) equal probability
from this ensemble.

Third, the statistic of interest is computed in the random bipartite network. The second and third steps are performed repeatedly, yielding a distribution of the statistic of interest observed in a set of random bipartite networks (i.e. under the null model). The set of random bipartite networks is known as an ensemble, and each randomly generated network can be viewed as a random sample from this ensemble. 

Finally, the statistic of interest from the observed network is compared to its distribution under the null model. Of particular interest is the proportion of times the statistic of interest under the null model is greater than or equal to the statistic of interest from the observed network. For example, observing a proportion of 0.02 would indicate that only 2\% of the random networks produced a statistic of interest that was larger than that from the observed network. This proportion is known as a $p$-value, and can be used in hypothesis testing concerning the randomness of the pattern captures by the statistic. In this example, using a conventional threshold of statistical significance, because $p < 0.05$, one would reject the null hypothesis that the pattern measured by the statistic of interest is random, and would instead conclude that the pattern is non-random.

In some cases, depending on the target metrics and on the chosen null model constraints, equivalent results could be achieved by analytical computation, when the randomization procedure provides a closed formula for the probability of existence of any link in the network. The analytical approach is called ``canonical'', while the randomization approach based on ``sampling'' random matrices is called ``microcanonical'' (see section~\ref{sec:how}). The canonical approach offers a clear advantage on the microcanonical one in that it eliminate the computational demand associated to matrix randomization techniques. However, there are various situations where only the microcanonical approach is possible.

\section{Bipartite null models}
Detecting patterns in bipartite networks involves comparing an observed network to an ensemble of random networks. However, there are multiple ways to conceptualize `random', with each conceptualization defining a slightly different ensemble and null model. In this section, we focus on null model choice and describe two features of the null model that are particularly important: which characteristics of the observed network should be preserved in the ensemble of random networks, and how this ensemble should be generated.

\subsection{What characteristics should be preserved?}
\label{sec:what}
Bipartite null models are primarily defined by which characteristics of the original bipartite network are contained in each randomly generated bipartite network. In principle, it is possible to imagine a null model that does not contain any of the original network’s characteristic. In this case, one generates an ensemble of bipartite networks, randomly drawn from a set of all existing bipartite networks. For practical but also scientific reasons it makes more sense to generate an ensemble as a subset of bipartite networks which contain some characteristics of the original one, i.e. the dimension (number of nodes), the fill (number of edges), and/or the marginal (node degrees).

First, all null models constrain the random network's dimensions. That is, they require that each random bipartite network under the null model has the same dimensions (i.e. the same number of each type of node) as the original. For example, if the original network's matrix representation has 5 rows and 10 columns, then all random networks generated under a null model will also have 5 rows and 10 columns.

Second, most null models also constrain the random network's fill. That is, they require that each random bipartite network under the null model contains the same number of 1s (i.e. the same number of edges) as the original. For example, if the original network's matrix representation contains 10 cells filled with a 1, then all random networks generated under a null model will also contain 10 cells filled with a 1.

Third, null models can vary in the constraints they impose on the random network’s row and column marginals (i.e. the degrees of the row and column nodes). Marginals can be unconstrained, such that the marginals in the randomly generated networks do not necessarily match those in the original. Or, they can be constrained ``softly'', such that the marginals in individual random networks do not necessarily match those in the original, but the average marginals of all the bipartite networks in the ensemble do. Or, they can be constrained exactly, such that the marginals in each random network always match those in the original (note that constraining exactly the marginal totals results in constraining exactly also matrix fill). Moreover, no, soft and exact constraints can be imposed on row marginals independently from column marginals.

\begin{figure}[!htbp]
	\begin{center}
		\includegraphics[width=1\textwidth]{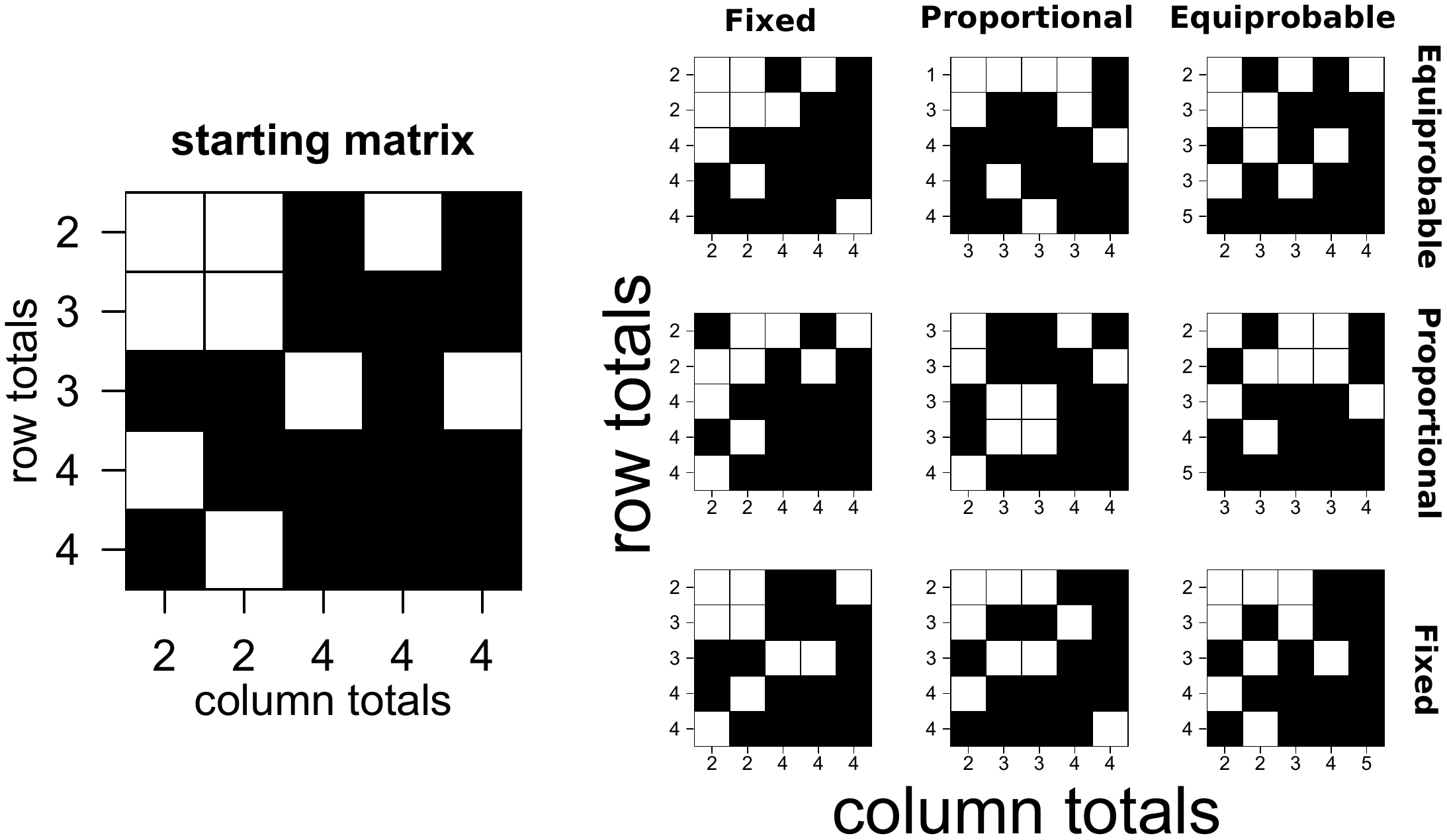}
	\end{center}
	\caption{A classification scheme of bipartite randomization algorithms, based on whether the matrix row and columns sums are preserved exactly ($Fixed$), preserved on average ($Proportional$), or unconstrained ($Equiprobable$). Black cells in each matrix indicate presence of links (i.e. $1$s) between the items in rows and the items in columns, while white cells indicate the absence of links (i.e. $0$s).}
	\label{fig3} 
\end{figure}

Figure \ref{fig3} illustrates how this generates different possible combinations defining different null models, some of which have specific names. For example, the highly constrained null model described by lower-left cell in the right panel of figure \ref{fig3} (which we can denote as ``Fixed-Fixed'', or ``FF''), requires that every random network has both row and column marginals that exactly match those in the original. This null model is sometimes called the Fixed Degree Sequence Model~\cite{zweig2011systematic}, and relies on an ensemble that is microcanonical. The somewhat less constrained null model in the central matrix of the right panel of figure \ref{fig3} (which we can denote as ``Proportional-Proportional'', or ``PP'') requires that the row and column marginals of the random networks only match those in the original \textit{on average}. This null model is sometimes called the Stochastic Degree Sequence Model~\cite{neal2014backbone} or Bipartite Configuration Model~\cite{saracco2015randomizing}, and relies on an ensemble that is canonical. While many other null models are possible, these two have become the most widely used.

\subsection{How are random bipartite networks generated?}
\label{sec:how}
Two broad approaches exist for generating random bipartite networks: fill methods and swap methods. 
Fill methods begin with an empty matrix with a fixed number of rows and columns
(fixed number of nodes for two modes), and incrementally add 0s and 1s as entries (edges between nodes). For example, the configuration model \cite{blanchet2013characterizing}  begins with an empty matrix, and has the condition to fill a certain fixed number of 1s in each row and column. The configuration model provides a way to generate a network
that satisfies the constraints of FF uniformly at random. In contrast, the
matrices that satisfy the constraints described by PP, where the column and row sums are fixed on average, can be generated by assigning each entry $M_ij$ a fixed probability. A Bernoulli trial for each entry gives then the decision if the entry is 1 or 0.

Swap methods begin with a network (often an observed bipartite network) and swap
the nodes (of one mode) from two randomly chosen edges, but only when the new possible edges have not already existed. In this case, a swap is not possible. In the matrix’ perspective, the algorithm starts with a matrix and swaps so called checkerboards (e.g., swapping $\begin{matrix}1&0\\0&1\end{matrix}$ with $\begin{matrix}0&1\\1&0\end{matrix}$). This approach was invented by Ryser in 1957\cite{ryser1957combinatorial}. More advanced
algorithms, known as Curveball algorithms, perform multiple swaps from two nodes (two rows) simultaneously \cite{strona2014fast,carstens2018unifying}. They were proven to be at least as efficient as the Ryser version\cite{carstens2015proof}. In practice, they perform often much more efficient than the classical version of Ryser\cite{carstens2017comparing}.

In the next section, we will show in detail how binary matrices can be randomized maintaining the constraints summarized in Fig.~\ref{fig3}. That section is intended for readers interested in getting a better understanding of the technical aspects behind matrix randomization, and could serve as a ``cookbook'' for a coding implementation of the various algorithms (or as a roadmap to help navigating the many implementations which are already available in several programming languages). Non interested readers can comfortably skip to section \ref{sec:choosing}.

\section{Randomization algorithms and procedures}
\subsection{Basic definitions and notation}
We denote a binary $(r,s)$-matrix as $M$, where $r$ is the number of rows and $s$ the number of columns. The size $S$ of $M$ is defined as $S=r \dot s$. We call the size of a binary matrix in some chapters also dimension. In two-mode networks, rows and columns correspond to distinct sets of real-world entities: for example, each row can be thought as representing an insect species while each column can be thought as representing a plant species.

The entry (or cell) $M_ij$ of matrix $M$ can have either value $0$ or $1$: $M_{ij} = 1$ indicates that the entity in the $i$-th row $r_i$ has some kind of association with the entity in the $j$-th column $c_j$ (for example, the $i$-th insect pollinates the $j$-th plant); $M_{ij} = 0$ indicates that no association exists (or has been observed) between the entities in the $i$-th row and $j$-th column. We will often refer to the $1$s as ``presences'', ``occurrences'' or ``filled cells'' (with identical meaning) and to the $0$s as ``absences'' or ``empty cells''. The total number of occurrences (i.e. the number of $1$s) in the $i$-th row and in the $j$-th column are denoted as $r_i = \sum_{j=1}^{c} M_{ij}$ and $c_j = \sum_{i=1}^{r} M_{ij}$ respectively. We will denote instances where $r_i = 0$ as ``empty rows'' and instances where $c_j = 0$ as ``empty columns''. Similarly, we will denote a matrix where all $M_{ij}$ entries are equal to $0$ an ``empty matrix''. We refer to the two sets of row and column totals as, respectively, $R = \{r_1\dots r_c\}$ and $C = \{c_1\dots c_r\}$. We denote the total number of occurrences in the matrix as $N = \sum_{i=1}^{r} r_i = \sum_{j=1}^{c} c_j $ and matrix fill as $\mathrm{fill} = N/S$. 

A binary matrix $M$ defined as above is equivalent to a bipartite network. In a bipartite network, we can identify two distinct sets of nodes which correspond to the two sets of real-world entities (e.g. plants and pollinators) identified by $M$'s rows and columns. Thus, the number of elements in the first set of nodes (e.g. plants) is equal to $r$ and the number of elements in the latter set of nodes (e.g. pollinators) is equal to $c$. 

The marginal totals of $M$ correspond to the so-called degrees of the nodes in the bipartite network. For instance, $r_i$ indicates the degree of the $i$-th node in the first set of nodes (e.g. the number of pollinators associated to the $i$-th plant), while $c_j$ corresponds to the degree of the $j$-th node in the second set of nodes (e.g. the number of plants associated to the $j$-th pollinator). Each entry $M_{ij} = 1$, correspond to an edge (or ``link'') in the bipartite network connecting the $i$-th node in the first set of nodes (e.g. a plant species) to the $j$-th node in the latter set (i.e. a pollinator species). Thus, $N$ will also correspond to the total number of edges in the bipartite network. In some scientific fields, the matrix $M$ is called the bi-adjacency matrix of the network.

We will refer to a single, randomized version of $M$ (that is, a new matrix obtained as the output of a given sampling algorithm) as $M^*$. For each randomization method, a set $\{M^*\}$ of randomized versions of $M$ are generated. Note that, in the statistical physics literature~\cite{Squartini2011a,squartini2015unbiased,squartini2017maximum,saracco2017inferring,cimini2019statistical}, the notation is usually the opposite, since the asterisk is used to denote the single empirical matrix, while the randomized matrices are left without an asterisk. The randomization algorithms we consider preserve, in three possible ways, the row and column sums of the original matrix $M$.

In the procedure that we denote as  $F$, the row and/or column sums of the real observed matrix are preserved exactly by randomization; in the procedure that we denote as $P$, the row and/or column sums in the randomized matrices match only on average (i.e. as an average over the generated set of randomized matrices) those of the original matrix; finally, in the procedure that we denote as $E$, the row and/or column sums of the randomized matrices are unconstrained and hence to a large extent  independent of those of the focal matrix. 

The concept of average constraints in the procedure $P$ has sometimes been misused in literature~\cite{bastolla2009architecture,strona2014methods}. To add more to the zoology of possible randomization algorithms, we note that some of them apply to the row and column sums different choices of the procedures $F$, $P$ and $E$. For this reason, we will describe the nature of the constraints of a given algorithm using the notation $XY$, where $X$ (respectively, $Y$) indicates the procedure applied to the row (respectively, column) sums of the original matrix. Both $X$ and $Y$ can take any of the three values $F$, $P$ and $E$, hence producing the 9 possible cases  illustrated in Fig~\ref{fig3}.

\subsection{Unconstrained rows, Unconstrained columns ($EE$)}

Method $EE$ is the most trivial method out of our 9 possible ones, which requires to sample matrices with fixed size $S$ and fill $f$ uniformly at random. This can be achieved by a variety of different approaches. In the class of swap methods, an efficient recipe is of exchanging entries $0$s and $1$s in the initial matrix in the following way. All entries $M_{ij}$ of matrix $M$ get a different number, starting from 1 to $S$ (the matrix size). These numbers can be permuted by a classical random permutation algorithm to get a new order of the numbers, and hence their corresponding entries $M_ij$. Let for example be a $(2,2)$-matrix with entries $M_{11}=0$ , $M_{12}=1$, $M_{21}=1$, $M_{22}=1$. $M_{11}$, $M_{12}$, $M_{21}$, $M_{22}$ get the numbers $1,2,3,4$ in this order. We permute the numbers in order $2,4,3,1$. Then the new matrix $M^*$ has the following entries: $M^*_{11}=1$, $M^*_{12}=1$, $M^*_{21}=1$, $M^*_{22}=0$. Classical random permutation algorithms come from mixing card decks and are known as random shuffles~\cite{Aldous1986SHUFFLINGCA}. They all have efficient running times. Note, that the number of all possible matrices of this class is ${S \over N}$.

$EE$ sampling can be also performed using filling approaches. There, we consider an empty $(r,s)$-matrix $M^*$, with $M^*_{ij}=0$ for all $i$ and $j$. Our goal is to fill that matrix with $N$ $1$s and $(S-N)$ $0$s. These values are taken from the initial matrix $M$. We give all entries $M^*_{ij}$ a different number from $1$ to $S$. $N$ times we use one after another a classical random number algorithm to first choose a number uniformly at random, and then delete it from the set of numbers. The corresponding entries $M^*_{ij}$ of all chosen numbers are set to $1$. The remaining entries to $0$. Classical random number algorithms are efficient and can be found for example in~\cite{knuth1998art}.

One important aspect to take into account is that both procedures might generate empty rows and/or columns. This might or might not be desirable/acceptable. If not, there are different potential solutions. In the first filling approach starting from an empty matrix, assuming $N\geq(r+c)$ one might first assign a $1$ to each entry $M_{ix}\:\forall\:i\in[1,r]$, with $x$ being a random integer in $[1\dots c]$, and then to each entry $M_{yj}\:\forall\:j\in[1,c]$, with $y$ being a random integer in $[1\dots r]$. In sparse matrices where $N<(r+c)$, then a different approach would be needed. Assuming that, for example, $r>c$ and $N\geq r$, one might first convert to $1$ all the $M_{ij}$ entries where $i=j$ and then attribute one presence in the each entry $M_{ix}\forall\:i\in[c,r]$ and $x$ being a random integer in $[1\dots c]$.

Alternatively, one can solve the Shannon-Gibbs entropy maximization problem, the only constraints being the normalization condition and the request $\langle N\rangle=\tilde{N}$; maximizing the corresponding likelihood leads to the expression $p_{ij}=N/S$. The canonical ensemble of the Bipartite Random Graph Model can be, then, explicitly sampled by considering each entry of $M$, drawing a real number $u_{ij}\in U[0,1]$ and posing $M_{ij}=1$ if $u_{ij}\leq p_{ij}$.

\subsection{Unconstrained rows, Fixed columns ($FE$ or $EF$)}
As for the $EE$ case, the $FE$ (and its equivalent $EF$) are quite trivial and can be easily and efficiently implemented using different approaches. Conceptually, $FE$ requires to randomize the position of $0$s and $1$s in each individual row of $M$. If only the position of presences and absences within a row is randomized but not their respective numbers, it is intuitive that $R$ in the randomized  matrix $M^*$ will remain the same as in $M$. In the jargon of physics, when the rectangular matrix represents the adjacency matrix of a bipartite network, this model is known as the Bipartite Partial Configuration Model (BiPCM) \cite{saracco2017inferring}, because the constraint is the degree of each node (`configuration model'), but it is enforced `partially', i.e. on only one of the two layers.

A simple algorithmic implementation of $FE$ might consist of generating $r$ random lists each including $r_i$ $1$s and $c-r_i$ $0$s, and then combine those lists into a matrix $M^*$. Similarly, for $EF$, one might generate $c$ random lists each including $c_j$ $1$s and $r-c_j$ $0$s, and then combine those lists into a matrix $M^*$.
Upon doing so, the microcanonical ensemble of the Bipartite Partial Configuration Model - whose cardinality amounts at $\Omega_\text{BiPCM}=\prod_{i=1}^R\binom{c}{r_i}$ - can be explicitly sampled.

Alternatively, one can solve the Shannon-Gibbs entropy maximization problem, the only constraints being the normalization condition and the `soft' request $\langle r_i\rangle=\tilde{r}_i$, $\forall\:i$; maximizing the corresponding likelihood leads to the expression $p_{ij}=r_i/c$. The canonical ensemble of the Bipartite Partial Configuration Model can be, then, explicitly sampled by considering each entry of $M$, drawing a uniform random real number $u_{ij}\in U[0,1]$ and posing $M_{ij}=1$ if $u_{ij}\leq p_{ij}$.
As in the case of $EE$, such approaches might result in generating empty columns or rows. If this is not desirable, one can implement additional steps/constraints in the randomization algorithms. For example, for $FE$, one might pre-assign a presence to the $j$-th position to a randomly selected row $\forall\:j\in[1,c]$. Then, the algorithm will be implemented as above, but presences and absences will be randomly placed in each row conditionally to the pre-assignments\footnote{The probability distribution of edge weights in $\mathbf{MM'}$ is given by the hypergeometric distribution \cite{neal2021comparing}.}.

\subsection{Constrained rows and columns (FF)}
This is the case that has received most of the attention from different fields, as relevant not only for a variety of practical applications but also for important theoretical questions in mathematics. For this reason, it has been the object of a large number of studies which have  produced a large corpus of methods which is constantly growing. As anticipated, such methods can be roughly classified into those which obtain a random matrix from scratch, i.e. by filling up at random an empty matrix, and those that randomize an existing matrix. We will first cover filling strategies and then move to randomization algorithms.
In the jargon of physics, this model is known with the name of microcanonical Bipartite Configuration Model (BiCM).

In principle, any \textbf{extant} random matrix $M$ of size $S = r \times c$ with row and column totals equals, respectively, to $R$ and $C$ can be obtained both by starting with an empty $r \times c$ matrix where each entry $M_{ij} = 0$ and then converting progressively entries to $1$ until the marginal totals matches exactly the expected $R$ and $C$. We emphasized the term ``extant'' as it is not for granted that an $r \times c$ matrix with $R=\{r_1\dots r_c\}$ and $C=\{c_1\dots c_r\}$ (let us call it $M(R,C)$) exists. Intuitively, an obvious necessary condition for the existence of the matrix is that the $\sum_{i=1}^{r}c_i=\sum_{j=1}^{c}r_j$. However, such necessary condition does not ensure the existence of $M(R,C)$. Indeed, the minimum necessary condition for the existence of $M(R,C)$ is provided by the 60 years Gale-Riser theorem \cite{gale1957theorem,ryser1957combinatorial}. If we follow the simple example provided by Gale \cite{gale1957theorem}, we can imagine that our matrix maps the placement of $r$ families going to a picnic across $c$ buses. There, $r_j$ is the total number of members in the $j$-th family, and $c_i$ is the total number of places available in the $i$-th bus. The theorem answers the question ``When is it possible to seat all passengers in such a way that no two members of the same family are in the same bus?''. Such a question is equivalent to asking whether it is theoretically possible to generate at least one $M(R,C)$ matrix.

The theorem provides the following necessary and sufficient condition for the existence of a solution to the problem:    

$\sum_{j=1}^{k} r_j \leq \sum_{j=1}^{k} s_j$,

for all integers $k$, where $s_j = \{c_i|c_i \geq j\}$, $c_i = 0$ for $i>c$ and $r_i=0$ for $i>r$ and with $c_i$ and $r_j$ being listed in decreasing order. 

The existence of the $M(R,C)$ matrix does not imply that generating it is an easy task. That is, if we start from matrix $M$ where each element $M_{ij} = 0$ and then we progressively modify randomly selected entries to $1$ checking at each step that the observed marginal totals do not exceed the expected $R$ and $C$ we will most likely end up in a situation where any further addition of a $1$ to $M_{ij}$ will lead to either exceed the expected $r_i$ or $c_j$. However, the minimum sufficient condition provide by the Gale-Riser problem offers also an efficient way to generate a matrix $M(R,C)$. In the example of the families and busses, if a solution exists, it will be always possible to succeed in placing all members of the different families in different busses (i.e. avoiding that two members of the same family are in the same bus) by allocating first all the members of the largest family to the busses having the largest number of available seats; then all the members of the second largest family to the bussess having most free seats after the allocation of the first family; then all the members of the third family to the bussess having most free seats after the allocation of the first and second family and so on. This procedure will always end with all persons seated, all members of each family seated in a different bus and no empty seats left in any bus. It is clear, however, that although this procedure permits to generate the $M(R,C)$ matrix, it will always generate an identical instance of all the possible matrices having marginal totals $R$ and $C$ while it is clear that we need some compromise permitting us to succeed in generating different $M(R,C)$ (i.e. without getting stuck in the allocation of $1$s to $M_{ij}$ entries before reaching the target $R$ and $C$) and to sample them with uniform probability from the universe of all possible $M(R,C)$ matrices.
  
Various approaches have been proposed for this purpose, but most of them have either problems in terms of computational efficiency or sample $M(R,C)$ matrices with different probabilities (or both). The ``knight tour'' algorithm proposed by Sanderson\cite{sanderson1998null} and its variations \cite{gotelli2001swap} try to fill progressively the matrix choosing randomly cells one at a time and ``backtracking'', i.e. returning to a previous state, when the procedure gets stuck, that is when it is no longer possible to fill-in a cell without exceeding $R$ or $C$. Besides being prone to biases \cite{gotelli2001swap}, these methods are impractical for even moderately sized matrices as the algorithm might spend a considerable (and hardly predictable) amount of time for backtracking \cite{gotelli2001swap}.
 
More recently, Chen et al.\cite{chen2005sequential} have proposed an approach based on ``sequential importance sampling'', which generates the matrix by sampling columns progressively. As noted by the authors, if the position of the $c_j$ $1$s of the $j$-th column is determined uniformly at random, it becomes extremely difficult to sample a valid column, which makes the process exceedingly computationally intensive. To overcome this issue, they proposed to generate the columns using the conditional-Poisson sampling method \cite{chen1994weighted, brewer2013sampling}, which, in a simplification, increases the chances to allocate a $1$ in the $i$-th position of the target $j$-th column if $r_i$ is large. This choice dramatically improves the computational efficiency of the method but prevent it from sampling matrices exactly from the uniform distribution, with the extent of the bias depending on both the actual setup of the conditional-Poisson sampling (i.e. the degree to which $r_i$ affects the probability of the $i-th$ element in the $j-th$ column to be a $1$) and the distribution of values in $R$ and $C$.

The alternative approach is that of using Markov Chain procedures where small incremental changes are applied to the target matrix. Those changes progressively bring the matrix far from its initial status. Ideally, if enough small changes are applied to the starting matrix, the probability of sampling any of all $M(R,C)$ matrices will converge to a uniform distribution. Clearly, the changes will need to ensure that the marginal totals of the initial matrix, $R$ and $C$, remain unaltered. The easiest--and most classical--way to achieve this goal consists in progressively selecting ``checkerboards'' and swapping their diagonal elements \cite{roberts1990island,gotelli2001swap}. A ``checkerboard'' is a specific pattern in the matrix where $M_{ij} = 0; M_{ik} = 1; M_{zj} = 1; M_{zk} = 0 $. It is intuitive that if we modify the matrix by ``swapping'' the diagonal elements of the checkerboard, that is by setting $M_{ij} = 1; M_{ik} = 0; M_{zj} = 0; M_{zk} = 1 $ the row ($r_i$ and $r_z$) and column ($c_j$ and $c_k$) totals will not change, leaving $R$ and $C$ unaltered. Note that the rows and columns forming the checkerboard do not need to be contiguous in the matrix. The move that is iteratively applied to the initial configuration in order to generate a family of randomized variants has been popularized with the name of local rewiring algorithm (LRA) in the literature concerning monopartite networks \cite{maslov2004detection,maslov2002specificity,milo2002network,stouffer2007evidence}.

One obvious drawback of this procedure is that each swap will produce a small modification in the matrix, so that a very large number of swaps will be required to generate ``truly random'' matrices (that is, matrices sampled at random from all possible ones). How many swaps ensure that each randomized matrix is sampled uniformly from the universe of possible $M(R,C)$ matrices is not clear - although a rule of thumb recommends it to be larger than $4N$, i.e. four times the total number of 1s in the network \cite{maslov2002specificity,milo2002network}. In practical implementations in the ecological field, the number of swaps used has been one or more orders of magnitude larger than the number of cells in the matrix. For instance, a common choice has been that of using $30$-$50$k swaps for matrices having size smaller than $100\times 100$ cells \cite{fayle2010reducing,gotelli2011over,fayle2011bias}.

The computational challenges emerging from the need of performing many swaps to generate an unbiased random matrix combined with that of generating many random matrices to perform robust tests in null model analysis has also generated the emergence of two different approaches in the implementation of swaps to randomize a matrix, namely sequential versus independent swap algorithms \cite{besag1989generalized,manly1995note,gotelli2001swap}. In the first class of algorithms, a pre-defined randomly chosen number of swaps (e.g. 30k) is applied to the original matrix to generate a single random matrix. In the latter class, an initial, large number of swaps is applied to the starting matrix to generate the first random matrix, while each subsequent random matrix is generated by applying a smaller number of swaps to the last generated matrix in the sequence. Clearly, the second approach is less computationally intensive when there is a need for generating a large set of random matrices. Still, biases in hypothesis testing might emerge from the potential non-independence of the random matrices in the sequence.

Recently, the computational challenges associated with classical swap algorithms have been partially overcome by more efficient approaches where the swaps are replaced by trade of elements between adjacency lists representing the set of neighbours of a focal node in the network representation of $M$ \cite{verhelst2008efficient,strona2014fast,godard2022fastball}. As in the original example provided by Strona et al. \cite{strona2014fast}, we can consider a matrix $M$ where the $r$ rows correspond to a set of kids, and the $c$ columns correspond to a set of different baseball cards. Each cell $M_{ij}$ in the matrix indicates whether ($1$) or not ($0$) the $i$-th kid owns the $j$-th card. Then, We can imagine that the kids meet during class break to trade cards and that trades happen according to the following two rules: (i) cards have identical value, meaning that one card is traded with exactly one card; (ii) kids are not interested in owning duplicated cards, thus a trade cannot take place if leading to such a situation. Now, a situation where, in compliance with the above rules, a kid trades a Babe Ruth with a Willie Mays, will correspond to a typical swap in the matrix. The number of cards owned by the two kids will remain the same, as well as the number of owners for the two cards. However, nothing prevents the two kids from trading more than one card. If we call $\{a\}$ and $\{b\}$ the sets of cards owned, respectively, by the first ($k_a$) and the second kid ($k_b$), we can identify the set of cards that $k_a$ can potentially trade with $k_b$ as $a_b = \{b\}-\{a\}$, and the set of cards that $k_b$ can potentially trade with $k_a$ as $b_a = \{a\}-\{b\}$. The two kids will be in a position to make a trade where $k_a$ gives $n$ cards sampled from $a_b$ to $k_b$ while receiving from $k_b$ and identical number of cards sampled from $b_a$, with $n$ being an integer varying between $0$ and the minimum of $a_b$ and $b_a$ sizes. Intuitively, as in the case of the single card trade, this exchange will result in no changes to the total number of cards respectively owned by $k_a$ and $k_b$, nor in the number of kids owning any of the traded cards.
           
The algorithmic implementation of such multiple card trades consist in first converting the matrix in a set of adjacency lists mapping the position of $1$s in each column for each row (or the position of $1$s in each row for each column). In the example above, such lists will include the set of cards owned by each kid (or the set of kids owning a certain card). Then, at each step, two lists will be drawn at random, and a trade of size $n$ (with $n$ being an integer randomly sampled with uniform probability between $0$ and the maximum number of tradable cards) will be performed. There are two distinct cases where a step will result in no changes in the underlying $M$, namely when the maximum number of tradable cards is $0$ (i.e. when $a_b$ or $b_a$ or both are empty) or when $n$ is randomly assigned a value of $0$. Carstens (2015) provided a formal proof that the Curveball algorithm is unbiased \cite{carstens2015proof}, that is it samples uniformly from the universe of all possible $M(R,C)$ matrices. However, she also showed that the algorithm remains unbiased even if ``no-trade shuffles'' are excluded from the Markov chain, that is if $n$ is sampled between 1 and the maximum number of tradable cards when the latter is $\geq1$. By contrast, she showed that if ``no-trade row pairs'', that is all the list comparisons where there are no tradable cards, are excluded, then the sampling is no longer guaranteed to be uniform.

By modifying larger portions of $M$ at each step, the Curveball and other similar algorithms \cite{verhelst2008efficient,godard2022fastball} speeds up dramatically the Markov Chain convergence with respect to older swap algorithms, that is they reach a virtually uniform sampling of $M(R,C)$ matrices in a much smaller number of steps \cite{strona2014fast,carstens2015proof,carstens2017comparing,carstens2018unifying}. However, both for the ``classical'' and the more recent approaches, how fast (that is, in how many steps) the selected algorithm converges towards the uniform sampling of random matrices for a given $M(R,C)$ is an open question.

Results from the network literature show that, for monopartite, undirected networks, uniformity holds (at least approximately) only when the degrees are such that $k_{max}\cdot{\overline{k^2}}/{(\overline{k})^2}$ is much smaller than the total number of nodes - with $k_{max}$ being the largest degree in the network, $\overline{k}$ being the average degree and $\overline{k^2}$ being the second moment~\cite{roberts2012unbiased}; for directed networks, a similar condition must hold). The violation of uniformity and ergodicity by the LRA implies that the quantities averaged over the graphs it generates are biased. In order to restore ergodicity, it is enough to introduce an additional ``triangular move'' inverting the direction of closed loops of three vertices; in order to restore uniformity, however, something much more complicated is needed: at each iteration, the attempted ``rewiring move'' must be accepted with a probability that depends on some complicated property of the current network configuration~\cite{roberts2012unbiased}. Since this property must be recalculated at each step, the resulting algorithm is extremely time consuming.

\subsection{Proportionally constrained rows and/or columns ($PP$,$PE$,$PF$)\label{PP-section}}
Randomization models that impose proportional constraints on both $\mathbf{M}$'s rows and columns aim to generate random matrices $\mathbf{M^*}$ such that the \textit{expected value} of $r_i\in\mathbf{M^*}=r_i\in\mathbf{M}$ and the \textit{expected value} of $c_j\in\mathbf{M^*}=c_j\in\mathbf{M}$. That is, for example, although $r_i$ in a specific randomly generated $\mathbf{M^*}$ may differ from $r_i\in\mathbf{M}$, the average value of $r_i$ across all $\mathbf{M^*}$ is equal to $r_i\in\mathbf{M}$. In the jargon of physics, this model is known with the name of canonical Bipartite Configuration Model \cite{saracco2015randomizing}.

Most $PP$ randomization models generate $\mathbf{M^*}$ via a filling approach in which $P(M^*_{ij})=p_{ij}$. That is, each cell in $\mathbf{M^*}$ is filled with either a $0$ or $1$ depending on the outcome of an independent Bernoulli trial with a $p_{ij}$ probability of success. Such models differ only in how they define $p_{ij}$\:\footnote{Probability distribution of edge weights in $\mathbf{MM'}$ is given by the Poisson-binomial distribution, where the parameters are derived from $p_{ij}$ \cite{neal2021comparing}.}. Valid values of $p_{ij}$ are subject to three constraints. First, because they must be well-defined probabilities, $0 \leq p_{ij} \leq 1$. Second, because the expected value of $r_i\in\mathbf{M^*} = r_i\in\mathbf{M}$, $\sum_{j=1}^{c} p_{ij} = r_i$. Third, because the expected value of $c_j\in\mathbf{M^*} = c_j\in\mathbf{M}$, $\sum_{i=1}^{r} p_{ij} = c_j$. However, within these constraints, many different values of $p_{ij}$ are possible. In practice, $PP$ randomization models aim to choose $p_{ij}$ such that $p_{ij} \approx P(M(R,C)_{ij} = 1)$.

Three types of methods exist for choosing $p_{ij}$ in this way. First, arithmetic can be used to define $p_{ij} = r_ic_j/N$, truncating out-of-bound values toward $0$ or $1$. This is the simplest and earliest approach \cite{gotelli2000null}. Second, fitted linear models can be used to define $p_{ij}$ as the value of $M_ij$ predicted as a function of $r_i$ and $c_j$ \cite{neal2014backbone}. We note that these Bernoulli trial approaches generate random matrices that do not retain the original marginal total distribution. This is because these trials follow a Poisson distribution that is nearly symmetrical at larger marginal values while being positively skewed at small values, by this causing higher simulated node numbers for comparatively low row or column total values.

Third, entropy maximization models can be used to define $p_{ij}$ \cite{neal2022homophily,saracco2017inferring}. In this case, the constraints defining the Shannon-Gibbs entropy maximization problem are the (usual) normalization condition and the request that $\langle r_i\rangle=\tilde{r}_i$, $\forall\:i$ and $\langle c_j\rangle=\tilde{c}_j$, $\forall\:j$: upon doing so, one finds an implicit expression for the probability coefficients $\{p_{ij}\}$ which is a function of the Lagrange multipliers employed to define the problem.

Although the numerical expression of the former ones can be determined only after the likelihood of the BiCM has been maximized, in case sparse matrices are considered, the following approximation\footnote{We would like to stress that the formula represents an approximation whose validity must be carefully checked before it can be safely employed: loosely speaking, networks must be sparse and hubs must be absent.} holds: $p_{ij}^\text{BiCM}=r_ic_j/N$. The canonical ensemble of the Bipartite Partial Configuration Model can, then, be explicitly sampled by considering each entry of $M$, drawing a real number $u_{ij}\in U[0,1]$ and posing $M_{ij}=1$ if $u_{ij}\leq p_{ij}$. In a recent review, \cite{neal2021comparing} found that the Bipartite Configuration Model was the fastest and most accurate method for computing $p_{ij}$.

Recently \cite{ulrich2012null} proposed another $PP$ randomization model that does not rely on probability-based cell filling. Rather than randomizing whether a cell is filled or not, this approach first randomly sets the target values for $r_i$ and $c_i$ in $\mathbf{M^*}$, then fills the cells of $\mathbf{M}$ to achieve these target values.

Algorithms where the proportional constraints are applied only to rows (or columns) in combination while the marginal totals of columns (or rows) are either kept fixed to $C$ or left to vary with equal probability (i.e. the $PF$ and $PE$ cases and their equivalents $FP$ and $EP$) are a theoretical possibility but have not received much attention and have been rarely used in real-world analyses. Gotelli \cite{gotelli2000null} provides some straightforward implementations. For the $PE$ model, one can start with an empty matrix and then add an occurrence to a cell $M^*ij$ with probability $r_i/(N \times c)$. For the $EP$ model, the probability of occupancy should be $c_i/(N \times r)$. For the $FP$ model, one might start with an empty matrix, and then add presences to rows one row at a time. For each $i$-th row, the algorithm reiterates the procedure of sampling a random $j$ with probability $c_j/N$ and setting $M_{ij}^*$ to $1$ until the total number of presences in the target row matches the expected value ($r_i$). For $PF$, for each $j$-th column, the algorithm reiterates the procedure of sampling a random $i$ with probability $r_i/N$ and setting $M_{ij}^*$ to $1$ until the total number of presences in the target column matches the expected value ($c_j$). Gotelli \cite{gotelli2000null} noted that the procedure can sometimes end up in matrices with either empty rows or columns, but do not provide a solution for that. Adjustments similar to those discussed for the $EF$ and $FE$ can be used to tackle this potential issue.

Notice that the Bipartite Configuration Model reduces to its partial versions (see section 5.2) once some of the constraints defining it are switched off \cite{saracco2017inferring}: more precisely, $PP$ reduces to $PE$ once the constraints concerning the columns are removed from the optimization problem; analogously, $PP$ reduces to $EP$ once the constraints concerning the rows are removed from the optimization problem.

\section{Choosing a null model}
\label{sec:choosing}
In the previous sections we have listed all the most common constraints/rules that can be taken into account to randomize a bipartite matrix, and we have described how such rules can be implemented into dedicated algorithms. However, we have not discussed a fundamental question which, even if it is not in itself central to the randomization procedures, constitutes the main reason for which these are developed. Why--or under which circumstances--one should choose one specific set of constraints over another?

There are many theoretical and methodological considerations that are relevant to choosing how to generate the random networks and which characteristics to preserve. The first aspect, is linked to computational and technical aspects. For a given randomization goal (such as generating a random matrix with the same marginal totals as the target one) there might be tradeoffs between the computational demand of a given algorithm, its reliability (for instance in terms of sampling uniformly from the universe of possible matrix configurations), and its easiness of implementation and integration in different analytical workflows. For instance, the recently introduced \textit{fastball} algorithm \cite{godard2022fastball} has a theoretical time complexity $O(n)$ time, while the earlier \textit{curveball} algorithm \cite{strona2014fast} has a theoretical time complexity of $O(n~log~n)$, but the practical running times of both algorithms depend on the programming language used to implement them. Since both algorithms permit sampling random matrices without biases, the choice of using one or another method depends on practical considerations related, among the others, to the actual amount of data to be processed and the coding integration with other analyses. If one has to randomize a few, small matrices, then using a simpler code implementation might be a reasonable choice, while for larger analyses the performance advantage might outweigh the potential additional effort in coding integration. 

However, it should be also noted that such additional effort might be in fact eased by the increasing availability of bindings between different programming languages (and especially between low and high level languages). Referring to the specific example reported here, the authors of the \textit{fastball} provided code which permits to run their algorithm in $R$ environment accessing directly compiled $C++$ code, hence combining the execution speed provided by the low level implementation with the user friendly syntax of the $R$ code. Analogously, a Python module that calculates BiCM probabilities at the speed of C was recently released~\cite{vallarano2021fast}. All this considered, in general we might assume that the choice of an implementation is possibly trivial, leaning most of the time towards the most efficient available tools. 

The opposite is true for choosing which constraints to apply to the randomized matrices. There the considerations revolve around the nature of the features the researcher wishes to control, and how those translate into specific matrix properties. A typical example is provided by the analysis of rectangular matrices representing the presence or absence of a set of plant or animal species across a set of localities (often islands or, in any case, isolated habitat patches). In that context, the marginal totals could be linked to different kind of ecological information. Specifically, for a matrix where rows correspond to species and columns correspond to islands, the column totals, representing species richness across localities, might be linked to various features affecting local species diversity. Some of these might be obvious and/or known, such as island size, while others, such as habitat heterogeneity or resource availability, or particular biogeographical features, might be less intuitive or not known. Still, one might assume that the effect of all of these features combined is actually reflected in column totals. Similarly, one could consider row totals, that is the prevalence of species across the islands, as a proxy for various features of the species, such as their ability to disperse and colonize islands and the generalism or specialization in their needs for resources.

Based on these considerations, one should then decide whether or not to preserve the marginal totals in the randomized matrices. To understand the choice, we need to make a clear distinction between ``patterns'', i.e. the different forms of organization of the various entities which are represented in the matrix and that are captured by ad-hoc metrics (such as nestedness \cite{patterson1986nested, patterson2000analyzing}); and the ``processes'' that led to the emergence of such patterns. Research questions usually target both patterns and processes, that is one could be interested in measuring whether and to what extent species are distributed across islands in a certain, non-random pattern, and what are the causes (i.e. processes) which led to such a pattern. However, these two objectives are not independent of one another. On the contrary they are two sides of the same coin.

To assess the relevance of a given process, one should ideally identify some way to isolate the effect of that process on the observed pattern from all the other processes that might be also involved in the emergence of the pattern. The null model approach which is central to this review offers one straightforward way to achieve this objective. In principle, one could explore the importance of a given process by comparing the target pattern in the original matrix with the same pattern in a large set of randomized matrices obtained by preserving all the features that might affect the emergence of the the observed pattern, with the exclusion of those features potentially emerging from the process of interest.
But this also means that the assessed magnitude of the pattern could vary depending on the identity of the target process. Thus, a matrix might show a strong structural pattern when examined with a focus on a given process, but no structure in a different context \cite{strona2018bi}.

Patterns in a matrix can be often described and measured by single values. For instance, one could measure the ``temperature'' (the original metric used to describe nestedness \cite{patterson1986nested}) of a given matrix and then use that only information to assess whether or not the matrix is structured, by placing the observed temperature within the theoretical range of possible values ($0-100$). However, as already emphasized multiple times within this review, such an approach might not be particularly enlightening. Specifically, most metrics of matrix structural patterns are not independent from matrix structural properties such as matrix size, shape, fill and marginal totals. Therefore, the metrics' raw values are a simultaneous result of the processes that determined the matrix properties and of \textit{other} processes. Such \textit{other} processes are usually central to interesting and meaningful questions, and standardizing the target metrics by controlling for matrix structural patterns is an obvious way to try isolating them. For example, by comparing the structure of a target species-island matrix with that of randomized versions having the same marginal totals one might be able to assess the structuring importance of some ecological processes other than those which determine local species richness and species prevalence across islands. Similarly, one could constrain selected structural properties to explore specific hypotheses or to answer specific questions. For instance, one could test the importance of local species richness in determining nestedness by comparing the target matrix with randomized versions obtained by constraining row marginal totals only (i.e. species prevalence across sites in our species-island matrix). 

To make a different example, we might consider a matrix mapping the authorship of scientific publications (with authors in rows and publications in columns). We can imagine a situation where one would be interested in quantifying the overall tendency for collaboration between authors. It is obvious that the frequency of co-authorship would naturally increase with the overall productivity of the scientific community represented by the matrix. However, it might be also reasonable to assume that the overall productivity is both a driver and a result of collaborations. Thus, it might be meaningful to assess the degree of co-authorship both taking or not taking into account the overall community productivity. One could also advance hypotheses on how the individual productivity of the different authors might affect the overall intensity of co-authorships (quantified by row totals in the matrix). That is, we might expect that a situation where all the authors have similar productivity would lead to different co-authorship patterns/levels compared to a situation where a few authors are highly productive while most authors are associated with few publications. A similar reasoning applies to the number of authors per publication (quantified by column totals in the matrix). Intuitively, we might expect different co-authorship patterns in a situation where most publications tend to have a similar number of co-authors compared to a situation where we have a few articles signed by many authors and most papers authored by few scientists. Again, depending on their actual goals, the investigators might decide to either constrain or not row and/or column totals when generating the randomized matrices to be used as a frame of reference to assess the target community's tendency for scientific collaboration.

Thus, in summary, there are no standard rules for choosing what constraints to apply when randomizing a given matrix. In ecological literature, there has been a tendency to support the use of strict constraints over weak constraints. The main reason for this is that, by using weak constraints (for instance, by generating random matrices preserving only matrix size and fill, but without setting any rule on marginal totals), one could easily overestimate the importance of a given pattern. If we compare a real-world dataset with a completely randomized one, we would have high chances that slight evidence will emerge strong, even if not particularly compelling in reality. By contrast, adding more constraints to the generation of the randomized matrices will permit us to perform a more balanced comparison between the real-world dataset and the randomized counterparts, and avoid to fall into type I errors (false positives). However, adding too many constraints might actually result in the opposite outcome. That is, it might lead to generating random matrices very similar to the starting one, and then to fail to detect any pattern from the comparison, even in situations where such pattern exist (type II errors, false negatives). 

\begin{figure}[!htbp]
	\vspace{20pt}
	\begin{center}
		\includegraphics[width=1\textwidth]{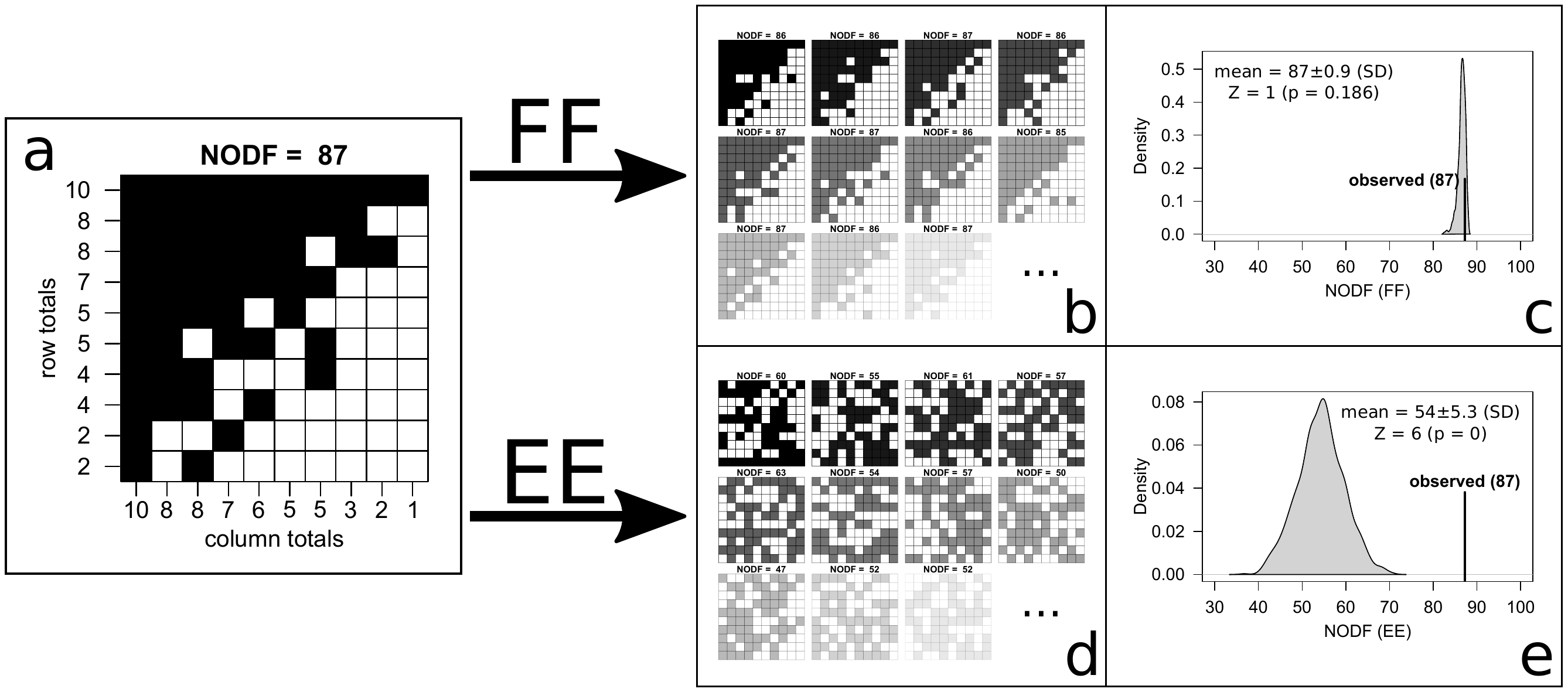}
	\end{center}
	\caption{\textbf{Effect of different randomization constraints on pattern detection.} Example of how applying different constraints to matrix randomization can lead to contrasting resulst in pattern detection. In this example we apply the same pattern detection workflow as described in Fig.~\ref{fig_pattern_analysis}. Black/grey cells in each matrix indicate presence of links (i.e. $1$s) between the items in rows and the items in columns, while white cells indicate the absence of links (i.e. $0$s). First, the structural measure of interest (in this case a nestedness metric, $NODF$ \cite{almeida2008consistent}) is computed on the target matrix (a). Then, two sets of 1000 randomized versions of the starting matrix are generated using, alternatively, an algorithm which generates random matrices with the same exact row and column totals of the starting matrix ($FF$); and an algorithm which generates random matrices having the same size, shape and fraction of occupied cells of the starting matrix, but with varying (equiprobable) row and column totals ($EE$). The target metric is computed for each random matrix in the two sets (b,d). Then, the starting $NODF$ value is compared against the two distribution of ``null'' values in the two sets of randomized matrices. In this example, the starting $NODF$ does not depart significantly from the null expectation from the set of matrices generated with the $FF$ algorithm ($Z = 1$; $p = 0.186$). Conversely, the pattern is identified as particularly strong when compared with the metrics measured in the random matrices generated with the $EE$ algorithm ($Z = 6$; $p = 0$).}
	\label{fig_choosing_constraints} 
\end{figure}

There is no silver bullet to solve this issue, with one workaround being that of exploring simultaneously a wide range of possible combinations of constraints and then placing and discussing the results within the multidimensional null modelling space identified by such constraints \cite{strona2018bi}. Each choice could be equally valid and useful to answer specific questions. In fact identifying such questions within a specific research context is a fundamental and dire challenge in itself, which should be regarded by researchers as a first--possibly the most important--step to be completed before even starting to think about the actual implementation of randomization routines. Unfortunately, this is not often the case, and often researchers make a fairly blind use of null models, without a clear reasoning behind the choice of randomization constraints. Overseeing the importance of linking solid questions about processes to null model pattern analysis might lead to a difficult or biased interpretation of the results. In the worst possible situation, one could even take advantage of the fact that different, sometimes contrasting results can arise when processing the same data using different randomization strategies (Fig.~\ref{fig_choosing_constraints}), and adjust procedural choices to steer the results in the desired direction. We hope that some of the information provided in this section and elsewhere in this review might help not only interested readers minimise the risk of falling into similar pitfalls when applying some of these techniques to their own analyses, but also that it might raise the attention of prospective reviewers on this important topic.

\section{Concluding remarks}
\label{sec:concluding}
In this review we have provided readers from different fields (such as, but not limited to, mathematics, physics, social sciences, ecology) with the conceptual tools needed to properly embark in matrix randomization exercises. Our hope is that it could also help unify future theoretical and applied research, and avoid the accumulation of further confusion due to duplicated efforts from different disciplines. However, there are various additional outstanding issues and open questions that we could not discuss here, but which we deem essential mentioning.

In most practical situations, and especially when referring to the natural world, detecting the existence of a link between two nodes (for instance by observing a pollinator's visit to a flower, or by detecting the presence of a parasite on a host) is much easier than quantifying the strength of the underlying association (e.g. the actual importance of the pollinator for the target plant, or the prevalence of the parasite species in the target host's population). Furthermore, different quantities could often be attributed to the same $0/1$ link, depending on the specific process which the target interaction represents. For instance, binary links connecting pollinators to plants might be associated to quantitative measures of pollinator preference, but also of pollination efficiency\cite{strona2022ecological}. As a consequence, there is a disproportion of studies--and tools--on binary matrices compared to quantitative ones. 

Nevertheless, although presence-absence matrices can be used to represent many different entities, zeros and ones cannot capture all the nuances and complexity which permeate the real world. Now, despite the challenges mentioned above, also thanks to novel tools and technologies, quantitative matrices are becoming increasingly available in many fields, and there is a growing recognition of the fact that weighting interactions might reveal different structural patterns from those identified after translating the same data into $0/1$ links \cite{staniczenko2013ghost}. The analysis of structural patterns in quantitative matrices might also require the use of randomization techniques. However, identifying a well defined set of criteria and constraints to be applied to the randomization procedures for quantitative matrices is not straightforward, and present many more possible cases than those identified for binary matrices (as in the classification scheme proposed in Figure \ref{fig3}). Since quantitative matrices underlie a binary structure (as each cell can be identified as either occupied or not), one could ideally perform randomizations by applying the same principles and techniques developed for $0/1$ matrices. However, in doing that, one should also decide whether to apply random changes to the individual values within each cell, or to preserve the original values while randomizing their position within the matrix, or to combine the two approaches. This opens up an extremely wide spectrum of possibilities, which becomes even wider when one starts thinking at possible alternative criteria and rules to modify (or not) the cell values \cite{ulrich2010null}. Remarkably, the canonical approach described in the text has already been straightforwardly extended to weighted bipartite networks, in the context of financial systems~\cite{di2018assessing}.

Another, obvious limitation of presence-absence or, more in general, of rectangular matrices, is that they can only capture a single feature of the system they are representing. That is to say, a rectangular matrix representing species occurrences across localities cannot provide any information on species and localities going beyond those we can directly derive from matrix structure. For instance, the matrix can tell us whether a given location has a high species richness, or whether a species is rare (clearly with specific reference to the set of localities included in the matrix). But it cannot tell us anything more about other ``exogenous'' features of species and localities. However, there are many possible contexts where such features not captured by the matrix itself could be relevant in the context of pattern detection. Such features might be used to define additional, external constraints. In addition to ``endogenous'' criteria for randomization looking at specific network/matrix properties, we might consider also ``exogenous'' criteria based on properties of the entities represented in the matrix which cannot be inferred by the matrix itself, but which can are made available by additional datasets. These can be, for example, additional data matrices with information relating to the rows or columns of the target matrix. This kind of information might be absolute, which is a simple  covariate or attribute of an individual component, or relative, which is a measure relating the component to other components. Considering again the species per locality matrix example, one could associate covariate vectors representing individual traits (body size, geographic range) to species, as well as relative measures such as the phylogenetic relationship of each species to all others. For sites, there might be physical covariates (soil nutrients, island area), but also relative measures such as the pairwise distances between all possible pairs of locations. The problem is particularly compelling in ecology, where it has led to the development of a few statistical approaches trying to improve pattern detection in species per site matrices by pairing them to additional information, incorporating, for instance, species functional traits and/or environmental characteristics of the sites \cite{ulrich2017comprehensive}. Although the implementation of exogenous constraints into ``typical'' randomization strategies might appear more as a conceptual challenge which one could tackle in practice by simply adding a few more lines to extant algorithms, the issue might be much more delicate. In fact, the additional constraints could modify fundamental properties of robust randomization techniques and lead to unexpected (and hardly detectable) biases. This calls for a more in-depth and formally structured investigation of the problem to identify potential extensions of extant, robust algorithms ideally capable of preserving the desired qualities of their original counterparts while accommodating extra rules dictated by features external to the target matrix.        

In this review, we focused on a set of different approaches to generate matrices under different constraints, such as that of ensuring that the randomized matrices generated by the selected approach have some pre-defined marginal totals. Such constraints, in principle, should permit a user's need to replicate the potential effects of real-world or hypothetical processes on matrix structure. For instance, by constraining column totals when applying a randomization algorithm to a species $\times$ locality matrix, one would generate random matrices where species richness in each locality matches that of the initial matrix. In turn, that would ideally make it possible to compare the structure of the starting matrix with that of the randomized matrices while controlling for all the processes affecting local species richness. However, depending on the research questions, the actual study setting and the nature of the data under investigation, the choice of which structural properties of the matrix to be controlled for in the randomization process might be not so obvious, and possibly biased by subjectivity. Additionally, one might be interested in exploring a gradient of assumptions and multiple scenarios. In such a context, a potential solution is that of comparing the target matrix not just with a specific set of null matrices obtained using a specific algorithm but, instead, with multiple sets of null matrices covering a larger--ideally continuous--portion of the null space entailed by different, specific combinations of constraint. For instance, in the ecological context, \cite{strona2018bi} have proposed an algorithm capable of exploring thoroughly the null space delimited by the 9 different combinations of constraints we focused on in this review (see fig.~\ref{fig3}). Such an algorithm can produce a ``bidimensional landscape'' of significance and effect size, in contrast with the typical single significance and effect size values provided by the standard null model analysis focusing on a specific set of constraints. The landscapes of significance/effect size, offer a more comprehensive and less subjective representation of matrix structural patterns, by showing how the intensity and significance of such patterns vary under a continuous range of different hypotheses.  

Finally, most of the existing algorithms suffer from the `curse of dimensionality', which make the algorithms hard to work on matrices with giant size. Designing scalable algorithms remains a challenging problem. Although the advances in high-performance and parallel computation might help tackling the challenge, novel technologies (such as next generation DNA techniques and remote sensing) are also generating datasets of increasing size. At the same time, novel questions are arising from the emergent data availability. This is to say, that the size of the matrices in need of randomization increases faster than the progresses in algorithmic optimization. The issue is further complicated by the fact that efficient parallel implementations are not available (and not easy to devise) for all the available randomization algorithms, although some important steps in this direction has been taken in the context of unimode network randomization \cite{carstens_et_al:LIPIcs:2018:9474}.

\textbf{Funding Acknowledgements}\\
ZPN was supported by USA National Science Foundation grants 2016320 and 2211744. DG and TS were supported by EU/NextGenerationEU/PNRR grant IR0000013. GW was supported by USA National Science Foundation grant 2210849. NJG was supported by USA National Science Foundation grant 2019470. STS was supported by a USA Air Force Office of Scientific Research grant FA9550-21-1-0140. WU was supported by an NCU institutional grant. Any opinions, findings, and conclusions or recommendations expressed in this material are those of the authors, and do not necessarily reflect the views of the funders.

\bibliographystyle{plain} 
\bibliography{references}

\begin{landscape}
	\begin{table}[]
		\caption{Examples of bipartite networks and their applications in the real world.}
		\label{table1}
		\begin{tabular}{llll}
			\hline
			\textbf{Field}  & \textbf{Subfield}                             & \textbf{System/application}          & \textbf{Example Reference (DOI)}       \\ \hline
			Biology         & Biomedical bipartite networks                 & Drug-target interaction networks     & \url{10.1093/bioinformatics/btaa157}   \\
			Biology         & Genetics                                      & Gene-sample binary mutation matrices & \url{10.1038/ng.3557; 10.1038/ng.3168} \\
			Biology         & Genetics                                      & Gene-tissue networks                 & \url{10.1101/517565 }                  \\
			Biology         & Molecular biology                             & Transcription factors                & \url{10.1126/science.aam8940}          \\
			Data Science    & Innovation                                    & Patents-technological codes networks & \url{10.1371/journal.pone.0230107}     \\
			Ecology         & Archeozoology                                 & Species-locality matrices            & \url{10.1016/j.jasrep.2015.02.008}     \\
			Ecology         & Ecological networks                           & Cleaning symbiosis                   & \url{10.1098/rsbl.2006.0562}           \\
			Ecology         & Ecological networks                           & Dung beetle-mammal networks          & \url{10.1111/ele.13095}                \\
			Ecology         & Ecological networks                           & Host-microbiota networks             & \url{10.1111/1365-2656.13297}          \\
			Ecology         & Ecological networks                           & Host-parasite networks               & \url{10.1038/ncomms12462}              \\
			Ecology         & Ecological networks                           & Plant-herbivore networks             & \url{10.1371/journal.pone.0052967}     \\
			Ecology         & Ecological networks                           & Mutualistic networks                 & \url{10.1111/j.1461-0248.2007.01061.x} \\
			Ecology         & Infectious disease ecology                    & Dbl$\alpha$ types vs isolates               & \url{10.1002/ece3.3803}                \\
			Ecology         & Meta-community ecology                        & Species-locality matrices            & \url{10.1007/BF00317508}               \\
			Economics       & E-commerce                                    & Buyers-products purchase networks    & \url{20.500.12469/2899}                \\
			Economics       & World trade web                               & Country-commodity networks           & \url{10.1038/srep10595}                \\
			Finance         & Cryptocurrency                                & Bitcoin lightning network            & \url{10.1088/1367-2630/aba062}         \\
			Finance         & Interbank networks                            & Exposure (borrower-lender) networks  & \url{10.1038/srep03357}                \\
			Mathematics     & Combinatorics/graph theory/matrix theory      & Network enumeration                  & \url{10.1016/0890-5401(89)90067-9}     \\
			Physics         & Social/ecological/economic/financial networks & Network reconstruction               & \url{10.1088/1367-2630/16/4/043022}    \\
			Physics         & Transportation networks                       & Bus route – bus stop networks        & \url{10.1016/j.physa.2006.10.071}      \\
			Physics         & Transportation networks                       & Ferry vehicle-flight networks        & \url{10.1016/j.omega.2019.102178}      \\
			Social Sciences & Blogger communities                           & Users-posts networks                 & \url{10.1016/j.physa.2012.06.004}      \\
			Social Sciences & Human behavior                                & Listeners-music groups               & \url{10.1103/PhysRevE.72.066107}       \\
			Social Sciences & Contact networks                              & Event-participation networks         & \url{10.1371/journal.pone.0171565}     \\ 
			 Social Sciences & Political science    &
			 Bill co-sponsorship & 
			 \url{10.2478/connections-2019.026}     \\ 
			\hline
		\end{tabular}
	\end{table}
\end{landscape}

\end{document}